\documentclass[runningheads]{llncs}

\usepackage{amssymb,amsmath}
\usepackage{graphicx}
\usepackage{url}
\usepackage{subfigure}
\urldef{\mailsa}\path|{Shawyanghua,
wentaowu1984}@gmail.com,{weiwang1,zhenying}@fudan.edu.cn|

\begin{document}

\mainmatter  % start of an individual contribution

\title{Efficient Algorithms for Node Disjoint Subgraph Homeomorphism Determination}

\titlerunning{Efficient Algorithms for ndSHD}

\author{Yanghua Xiao, Wentao Wu, Wei Wang and Zhengying He}
%\author{}
%
%\authorrunning{Efficient Algorithms for Node Disjoint Subgraph Homeomorphism Determination}

% the affiliations are given next
\institute{Department of Computing and Information Technology£¬\\ FuDan University,  ShangHai, China\\
\mailsa\\}
%\institute{}

%\toctitle{Efficient Algorithms for Node Disjoint Subgraph
%Homeomorphism Determination}
\maketitle

\begin{abstract}
Recently, great efforts have been dedicated to researches on the
management of large scale graph based data such as WWW, social
networks, biological networks. In the study of graph based data
management, node disjoint subgraph homeomorphism relation between
graphs is more suitable than (sub)graph isomorphism in many cases,
especially in those cases that node skipping and node mismatching
are allowed. However, no efficient node disjoint subgraph
homeomorphism determination (ndSHD) algorithms have been available.
In this paper, we propose two computationally efficient ndSHD
algorithms based on state spaces searching with backtracking, which
employ many heuristics to prune the search spaces. Experimental
results on synthetic data sets show that the proposed algorithms are
efficient, require relative little time in most of the testing
cases, can scale to large or dense graphs, and can accommodate to
more complex fuzzy matching cases.

\end{abstract}

\small
\section{Introduction}

Recently, large scale graph based data management has received more
and more research attentions, due to the broad application of graph
based data. In the study of graph based data management, \emph{graph
based pattern matching}, i.e., to determine whether the structure of
a pattern graph can match to that of a data graph, is the key of
many problems about graph data management.

Existing graph pattern matchings can be classified into two
preliminary categories: \emph{exact matching} and \emph{inexact
matching}. Exact matching requires that the matched two graphs are
isomorphic to each other; i.e., exact graph pattern matching is
based on \emph{graph isomorphism} relations between graphs. While
the inexact graph matching is often considered as \emph{subgraph
isomorphism} between graphs, which means that pattern graph $P$
matches to data graph $G$ if and only if $P$ is subgraph isomorphic
to $G$.

\begin{figure}
\begin{minipage}[t]{0.5\linewidth}
\centering
\includegraphics[width=1.7in]{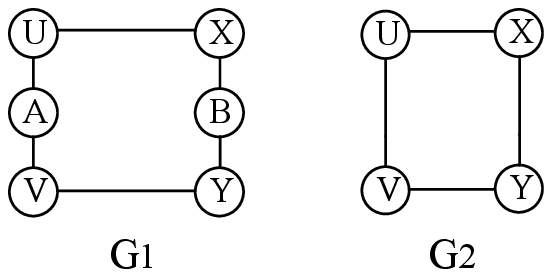}
\caption{Inexact Matching} \label{fig:inexact_match}
\end{minipage}
\begin{minipage}[t]{0.5\linewidth}
\centering
\includegraphics[width=2.5in]{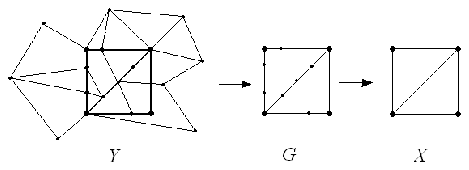}
\caption{Topological Minor} \label{fig:minor}
\end{minipage}
\end{figure}

However, in real applications, inexact graph pattern matching based
on subgraph isomorphism cannot represent the fuzzy matching in some
cases that \emph{node skipping or node mismatching is allowed.} For
example, as shown in Figure \ref{fig:inexact_match}, although
$G_{2}$ is not a subgraph of $G_{1}$, $G_{2}$ still can be regarded
as matched to $G_{1}$ if node skipping or node mismatching is
allowed. In other words, $G_{2}$ is matched to $G_{1}$ from the
abstract topological structure perspective, because $G_{2}$ retains
the abstract topological structure of $G_{1}$ if paths in $G_{1}$
can be contracted into the corresponding edges in $G_{2}$.

However, this kind of fuzzy matching is more desired in many real
applications than subgraph isomorphism based inexact matching. For
instance, the discovery of frequent conserved subgraph patterns from
protein interaction networks \cite{ref9,ref10} is an important and
challenging work in evolutionary and comparative biology, where
'conserved' just means the inexact graph pattern matching allowing
node mismatch and node skipping.  Similarly, in social network
analysis, the direct connection between nodes usually is not the
focus; instead, the high-level topological structure with
independent paths contracted is of great interest.

Using \emph{Graph Minor} theory \cite{ref2}, the abstract
topological structure in many real applications can be described as
\emph{topological minor}, and the relation between abstract
topological structure and its detailed original graph can be
described as \emph{node/vertex disjoint subgraph homeomorphism}.
However, to determine whether a pattern graph $P$ is a topological
minor of data graph $G$ is not a trivial thing, and this problem has
been proved to be NP-complete when $P$ and $G$ are not fixed
\cite{ref0}. Although Robertson and Seymour \cite{ref2} have
proposed a framework to solve \emph{minor containment} problem that
is a generalization of topology containment problem and \cite{ref3}
has implemented the framework, no efficient algorithms have been
dedicated to solve \emph{ndSHD} (in other contexts, also known as
\emph{topological minor containment}, \emph{homeomorphic embedding}
or \emph{topological embedding}), to the best of our knowledge.

To efficiently determine the node disjoint homeomorphism relation
between two graphs, we propose two algorithms based on state space
searching with backtrack, which integrate many heuristics into the
searching procedure to prune the search spaces. The work in the
paper is inspired by Ullmann's \cite{ref4} \b{s}ubgraph
\b{i}somorphism \b{d}etermination (SID) algorithm. However, for
ndSHD, we need to do some more specific things. First, for ndSHD,
not only \emph{node mapping space} but also \emph{edge-path mapping
space} needs to be searched, whereas for SID only the former needs
to be searched. Second, for ndSHD, according to the definition of
topological minor, we need to perform pairwise independence
determination of the paths to ensure the paths are disjoint. Third,
for SID, only edge information is explored, while in ndSHD path
information is explored too, which will be a great challenge to the
efficiency of the algorithm since the amount of paths is exponential
to the size of the graph.

In a summary, we make the following contributions in this paper:
\begin{enumerate}
\item We propose two efficient algorithms for node disjoint subgraph homeomorphism
determination. To the best of our knowledge, it's the first paper
dedicated to design practical efficient algorithms for node disjoint
subgraph homeomorphism determination or topological minor
containment determination problem.
\item We investigate the properties of topological minors, and employ these properties
as the heuristics to prune the search space.
\item We present a systematic performance study of proposed algorithms. The experimental results
show that the algorithms are efficient and scalable on synthetic
data sets.
\end{enumerate}

\section{Preliminaries}

We begin with some basic notations that are used in \cite{ref5}. Let
$G= (V,E,l)$ be a \emph{vertex labeled graph}, where $V$ is the set
of vertices, $E$ is the set of edges and $E\subseteq V\times V$, and
$l$ is a label function $l:V\rightarrow L$ , giving every vertex a
label.(In this paper, we only focus on vertex labeled graphs.
Unlabeled graph can be considered as a labeled graph with all
vertexes having the same vertex label.) The vertex set of $G$ is
referred to as $V(G)$, and its edges set as $E(G)$. A \emph{path}
$P$ in a graph is a sequence of vertices
$v_{1}$,$v_{2}$,...,$v_{k}$, where $v_{i}\in V$ and $v_{i}v_{i+1}\in
E$. The vertices $v_{1}$ and $v_{k}$ are linked by $P$ and are
called its \emph{ends}. The number of edges of a path is its
\emph{length}, and the path of length $k$ is denoted as $P^{k}$. A
path is \emph{simple} if its vertices are all distinct.
Particularly, a group of paths are \emph{independent} if none of the
paths have an inner vertex on another path. In the other words, a
path intersecting with other paths only at its ends can be called as
an \emph{independent path}. Be aware that the independent paths are
the key to study topological minors of a graph.

\subsection{Topological Minor}

As described in \cite{ref5}, a topological minor of a graph is
obtained by contracting the independent paths of one of its
subgraphs into edges. For example, in Figure \ref{fig:minor}, $X$ is
a topological minor of $Y$, since $X$ can be obtained by contracting
the independent paths of $G$ which is a subgraph of $Y$. Clearly,
contracting independent paths helps simplify a (sub)graph without
compromising its abstract topological information.

Formally, as shown in Figure 2, if we replace all the edges of $X$
with independent paths between their ends, so that these paths are
\emph{pairwise node independent}, namely none of these paths has an
inner vertex on another path, then $G$ is a \emph{subdivision} of
$X$, denoted as $T(X)$. If $G$ is a subgraph of $Y$, then $X$ is a
\emph{topological minor} of $Y$. As a subdivision of $X$ and a
subgraph of $Y$, if $G$ is obtained by replacing all the edges of
$X$ with independent paths with length from $l$ to $h$, then $G$ is
a \emph{ (l, h)-subdivision} of $X$ and $T$ is a \emph{(l,
h)-topological minor} of $Y$.

Given two graph $X$ and $Y$, if $X$ is a topological minor of $Y$,
then there exists a corresponding \emph{node disjoint subgraph
homeomorphism} from $X$ into $Y$, which
 is a pair of injective mappings $(f, g)$ from $X$ into $Y$,  where $f$ is an injective mapping from vertex set of
 $X$ into that of $Y$ and $g$ is an injective mapping from edges of $X$ into simple paths of
$Y$ such that (1) for each $e(v_1,v_2)\in E(X)$, $g(e)$ is a simple
path in $Y$ with $f(v_1)$ and $f(v_2)$ as two ends;(2) all mapped
paths are pairwise independent. In other words, if $X$ is node
disjoint subgraph homeomorphic to $Y$, all the edges of $X$ can be
mapped to a corresponding simple path of $Y$ and all the mapped path
are pairwise independent; all the nodes in $X$ can be mapped to a
corresponding node in $Y$(all the mapped nodes are called
\emph{branch nodes} of $Y$).

\subsection{Problem Definition}

\begin{figure}
\begin{minipage}[t]{0.4\linewidth}
\centering
\includegraphics[width=1.8in]{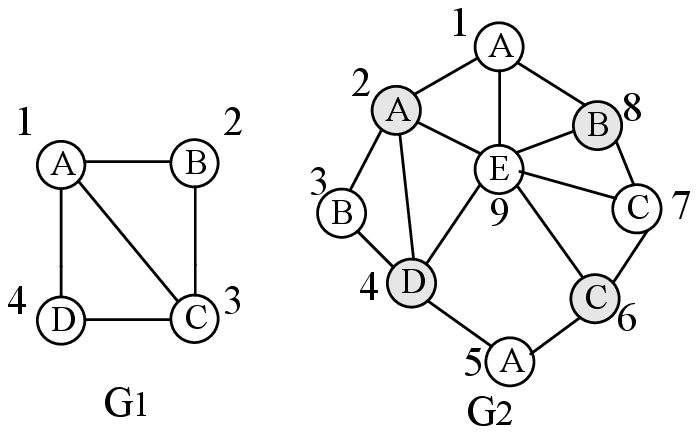}
\caption{Running Example} \label{fig:example}
\end{minipage}
\begin{minipage}[t]{0.6\linewidth}
\centering
\includegraphics[width=3in]{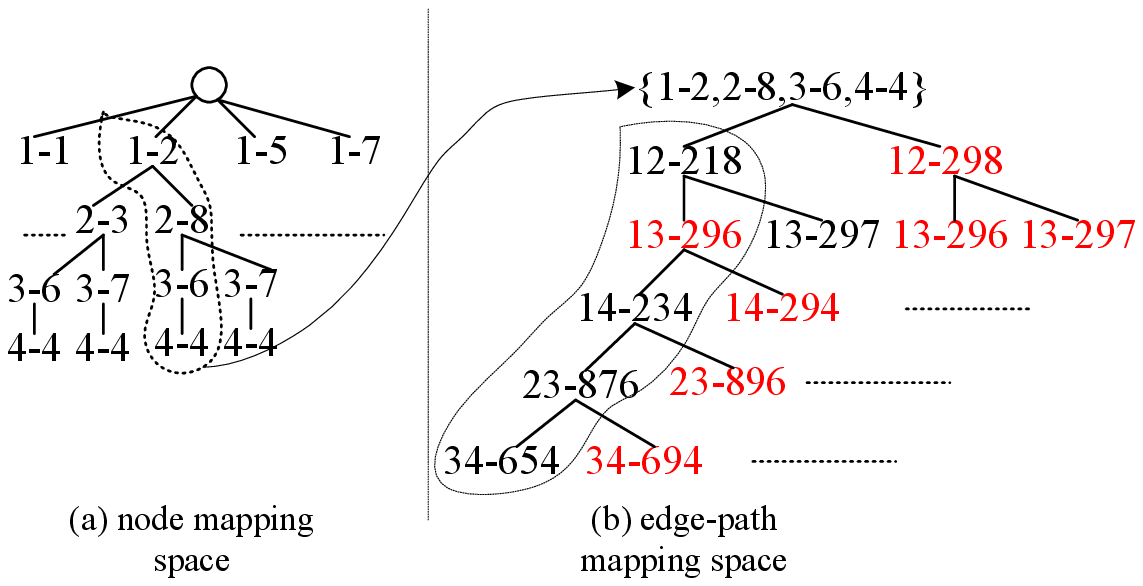}
\caption{Two Level State Space Searching} \label{fig:run_example}
\end{minipage}
\end{figure}

As shown in Figure \ref{fig:example}, given two vertex labeled
graphs $G_{1}$ and $G_{2}$, given the minimal path length $l$ and
the maximal path length $h$, the problem is whether $G_{1}$ is a
$(l, h)$-topological minor of $G_{2}$, i.e., $G_{1}$ is node
disjoint homeomorphic to $G_{2}$ so that all mapped paths in $G_{2}$
have length from $l$ to $h$. Obviously, this problem is a typical
\emph{determination problem}. When the answer is true, the
homeomorphism mapping $(f, g)$ also can be obtained. The solution to
the determination problem also can be extended to solve the
\emph{enumeration problem}, which is to find the entire valid
homeomorphism mappings between two graphs.

The answer to the problem is sensitive to the given parameter $(l,
h)$. For example, in Figure 3, if $(l, h)$ is $(2,2)$, which means
the edges in $G_{1}$ can only be mapped to the paths in $G_{2}$ with
length $2$, then nodes in $G_{1}$ can be mapped to the four nodes in
shadow in $G_{2}$ and the five edge-path mappings are ${12-218,
13-296, 14-234, 23-876, 34-654}$. If $(l, h)$ is $(3,3)$, $G_{1}$ is
not a topological minor of $G_{2}$. The influence of parameter $(l,
h)$ on topology containment determination has been discussed in
\cite{ref6} in detail.

\section{Algorithm Framework}

To simplify the description of the algorithm, we first give some
notations. Assume that vertex labeled graph
$G_{1}=(V_{1},E_{1},l_{1})$ is a $(l, h)$-topological minor of
vertex labeled graph $G_{2}=(V_{2},E_{2},l_{2})$ under the node
disjoint subgraph homeomorphism $(f,g)$, where $f:V_{1}\rightarrow
V_{2}$ and $g:E_{1}\rightarrow P^{l}\cup ...\cup P^{h}$, the image
of $E_{1}$ under mapping $g$ is denoted as $g(E_{1})=\{g(e)|e\in
E_{1}\}$. The number of vertices and edges of $G_{1}$ and $G_{2}$
are $n_{1}$, $m_{1}$ and $n_{2}$, $m_{2}$, respectively. For the
convenience of notation, we call $G_1$ as \emph{minor graph}, and
$G_2$ as data graph; without explicit statement, in the following
discussion, $G_1$ always denote a minor graph, $G_2$ always denote a
data graph.

\subsection{A Rudimentary Algorithm}

To determine whether $G_{1}$ is a $(l, h)$ topological minor of
$G_{2}$ is equivalent to find a pair of mapping $(f, g)$ between
these two graphs. The mapping $f$ maps the nodes in $G_{1}$ to the
nodes with the same label in $G_{2}$ so that $g$ can map each edge
of $G_{1}$ to a corresponding path in $G_{2}$. Obviously, the final
solution of the determination, i.e., the complete mapping $(f, g)$
between these two graphs, can be described as
$\mathcal{M}=(NM,EPM)$,where $NM\subseteq V_{1}\times V_{2}$ is the
node match set and $EPM\subseteq E_{1}\times(P^{l}\cup...\cup
P^{h})$ is the edge-path match set. All the mapped nodes of $G_{2}$
can be denoted as $NM^{(2)}$, and all the mapped paths of $G_{2}$
can be denoted as $EPM^{(2)}$.

The process of finding the homeomorphism mapping can be suitably
described by means of \emph{State Space Representation} \cite{ref7}.
Each state $s$ of the matching process can be associated with a
partial mapping solution $\mathcal{M}_s=(NM_s,EPM_s)$, where $NM_s$
and $EPM_s$ are the node match set and edge-path match set at state
$s$, respectively. Obviously, $\mathcal{M}_s$ contains all the
matches we have found so far and probably become a subset of some
final match set $\mathcal{M}$.

Given the two vertex labeled graphs as shown in Figure 3, a naive
two level state space searching procedure for a $(2,2)$ topological
mapping is shown as Figure 4, where the first level is to find a
suitable node mapping solution (shown in the dotted box of Figure
\ref{fig:run_example}(a)) and the second level is to find a suitable
edge-path mapping solution (shown in the dotted box of Figure
\ref{fig:run_example}(b)). The corresponding algorithm framework is
shown as follows.

\medskip
\noindent{\textbf{Algorithm}
ndSHD1($G_{1}$,$G_{2}$,$l$,$h$)}
\newline \noindent{\small
\textbf{Input:} $G_{1}$,$G_{2}$:vertex labeled graphs; $l$:minimal
path length; $h$:maximal path length.}\newline \noindent{\small
\textbf{Output:} If $G_{1}$ is a $(l,h)$-topological minor of
$G_{2}$ return \emph{true} and return \emph{the \textbf{first} found
node disjoint subgraph homeomorphism} $(f,g)$, otherwise return
\emph{false}.}

\small\begin{enumerate}
\item \small Initial($M$);/*Initialize SHD, Generate necessary path
information, Initialize the basic data structures*/

\item \small Initial($R$);
\item \small $s\leftarrow\emptyset$; /*initialize state as empty state*/
\item \small $s\leftarrow$NodeMappingSearch($s$,$M$,$R$); /*node mapping space search*/
\item \small \textbf{if not} IsValid($s$)
\item \small \hspace{2mm} \textbf{return false};
\item \small \textbf{else}
\item \small \hspace{2mm} $s\leftarrow$EdgePathMappingSearch($s$); /*edge-path mapping space*/
\item \small \textbf{if not} IsValid($s$)
\item \small \hspace{2mm} \textbf{return false};
\item \small \textbf{else}
\item \small \hspace{2mm} \textbf{return true};
\end{enumerate}

At first, we initialize two basic data structures : \emph{node
compatible matrix} $M$ and \emph{independent path matrix} $R$ as
well as its associated \emph{path indexed structure}. Then we start
the node matching process from the empty state. Each time we select
a branch in the state space, a state $s$ transits to a new successor
state $s$' by adding a new match, which is a node pair or an edge
path pair, to the partial solution. Each time a new match state
arrives, $M$ and $R$ are updated so that the node mapping space and
edge-path mapping space can be pruned. When a complete node mapping
has been found, the matching process will come to the second level:
edge-path matching space search. Similar to the search process in
node mapping space, each time a branch is selected, an edge-path
pair is added to the partial mapping solution and the independent
path matrix is updated. The process continues until a complete
edge-path mapping is found.

In the above searching process, if all the possible valid branches
in the subspace rooted at current state $s$ have been explored, but
still no valid match can be found, the searching process
\emph{backtracks} to the parent state of $s$. And any time the
procedure enters into \emph{dead state} which will be discussed in
3.5, the whole process will stop and return false which means the
two graphs do not satisfy the $(l,h)$-topological minor
relationship.

\subsection{Basic Data Structure}

As described above, we need to two basic data structures, one is
used to represent the node mapping information; the other is used to
represent $(l,h)$ independent path information of $G_{2}$. For the
former, we use node compatible matrix; the latter, we use
independent path matrix as well as a path index structure. Both of
them are changing with the transition of the matching state.

\begin{figure}
\begin{minipage}[t]{0.4\linewidth}
\centering
\includegraphics[width=2in]{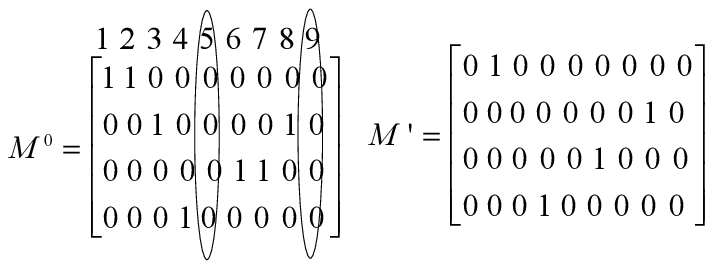}
\caption{$M^0$ and $M'$} \label{fig:side:a}
\end{minipage}
\begin{minipage}[t]{0.6\linewidth}
\centering
\includegraphics[width=2.2in]{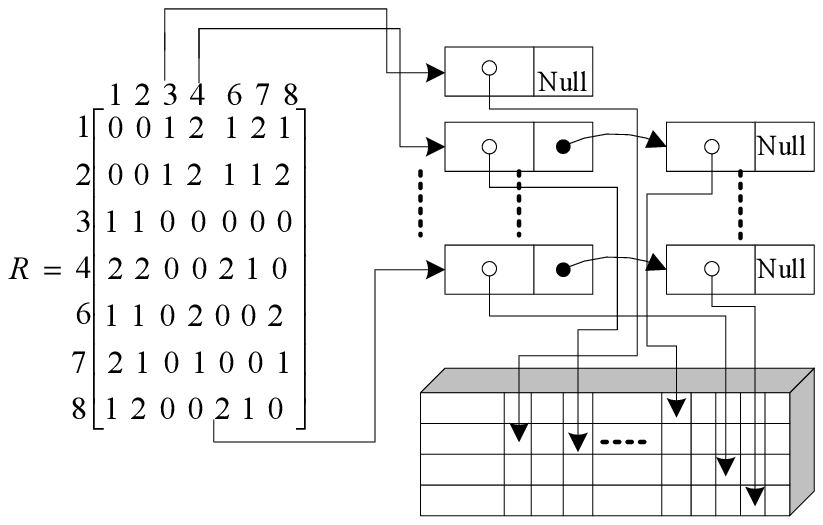}
\caption{R and its associated Path Indexed Structure }
\label{fig:side:a}
\end{minipage}
\end{figure}

We define node compatible matrix $M=[m_{ij}]$ to be a $n_{1}$
(rows)$\times n_{2}$ (columns) matrix whose elements are $1$'s or
$0$'s. At the final success state, we can get a final mapping matrix
$M'=[m'_{ij}]$ whose elements are $1$'s or $0$'s, such that each row
contains exactly one $1$ and each column contains no more than one
$1$. The final mapping matrix represents a valid one to one mapping
between nodes of $G_{1}$ and $G_{2}$, while the initial compatible
matrix $M^0$ represents the probable mappings between nodes of
$G_{1}$ and $G_{2}$. The initial node mapping and the final node
mapping between $G_{1}$ and $G_{2}$ in Figure 3 is shown in Figure
5. Obviously for each element $m'_{ij}$ of $M'$,
$(m'_{ij}=1)\rightarrow (m^{0}_{ij}=1)$.

Clearly, to reduce the number of $1$'s in $M$ is the key to speed up
the search procedure in node mapping space. Hence, the first key
step is to construct an initial compatible matrix $M^{0}$ with as
less $1$'s as possible. For this reason, we first introduce Lemma 1.
Due to the limitation of space, the detailed proof is omitted in
this paper.

\begin{lemma}
The number of elements of the independent path set starting from a
specified vertex $v_{i}\in V$ is no more than $d(v_{i})$, where
$d(v_{i})$ denotes the degree of $v_{i}$.
\end{lemma}

Since every path set starting from $v_{i}$ necessarily pass through
one or more edges incident with $v_{i}$, the independent path set
starting from $v_{i}$ has at most $d(v_{i})$ elements. According to
lemma 1, the node $v$ in $G_{1}$ cannot be matched to those nodes in
$G_{2}$ whose degree is less than $d(v)$. Therefore, we construct
the initial compatible matrix $M^{0}$ in accordance with the
following rule: $m^{0}_{ij}=1$ if $l_{1}(v_{i})=l_{2}(v_{j})\wedge
d(v_{i})\leq d(v_{j})$, otherwise $0$. As shown in Figure 3, $v_{1}$
in $G_{1}$ cannot mapped to $v_{5}$ of $G_{2}$, although these two
nodes have the same label.

When constructing Independent Path Matrix and its associated Path
Indexed Structure, the first problem we face is whether we need to
generate all the $(l, h)$ path information of $G_{2}$. The answer is
false, which is based on the following lemma.

\begin{lemma}
If $G_{1}$ is a $(l,h)$-topological minor of $G_{2}$ under subgraph
homeomorphism $(f,g)$, then $g(E_{1})$ only contains paths ending
with those branch nodes in $G_{2}$.
\end{lemma}

For example, in Figure 3, since $v_{5}$ in $G_{2}$ cannot be a
branch node, then all paths starting from $v_{5}$ needn't to be
enumerated. However, note that the path having $v_{5}$ as inner
vertex can not be ignored.

Therefore, we only need to enumerate all the $(l,h)$ paths between
all \emph{candidate branch node} pairs. These candidate branch nodes
can be filtered out by matrix $M^{0}$. As shown in Figure
\ref{fig:example}, since column $5$ and $9$ have only $0$'s, $v_{5}$
and $v_{9}$ in $G_{2}$ cannot be branch nodes, thus could be
filtered out and the remaining nodes in $G_{2}$ are just candidate
branch nodes. The cardinality of the candidate branch node set is
denoted as $n'_{2}$.

Then, we can define the independent path matrix $R=[r_{ij}]$ to be
$n'_{2}$(rows)$\times n'_{2}$ \newline(columns) matrix whose
elements are positive integers or $0$'s, which represent the number
of $(l,h)$ paths between the node pair $(v_{i},v_{j})$ in $G_{2}$.
The corresponding detailed path information are stored in a list
array \emph{RLists}, where each list in RLists contains
corresponding path addresses that point to the physical storage of
the path. RLists can be considered as a path index structure that is
built according to the end vertex pair of the path.

\subsection{State Space Searching}

The procedure of node mapping space searching and edge-path mapping
space searching are similar to each other. These two procedures are
shown as follows.

\medskip

\noindent{\textbf{Algorithm} Node/EdgePatbMappingSearch1
($s$,$M$,$R$)\newline \noindent{\small \textbf{Input:} $s$:the
current matching state; $M$:the current node compatible matrix; $R$:
the current independent path matrix.}\newline \noindent{\small
\textbf{Output:} \emph{found}: a boolean variable indicating whether
a complete node/edge-path mapping has been found.}

\begin{enumerate}
\item \small \textbf{if}($s$ is \emph{dead state})
\item \small \hspace{2mm} \textbf{return false};
\item \small \textbf{if}($s$ is \emph{complete mapping state})
\item \small \hspace{2mm} \textbf{return true};
\item \small let \emph{found}$\leftarrow$\textbf{false}
\item \small \textbf{while}(\textbf{not} \emph{found} \&\& Exists
Valid node/edge-path Mapping Pair)
\item \small \hspace{2mm} $m\leftarrow$GetNextNodePair();
/*$m\leftarrow$GetNextEdgePathPair();*/
\item \small \hspace{2mm} $s'\leftarrow$BackupState($s$);
\item \small \hspace{2mm} $NM_s\leftarrow NM_s\cup\{m\}$; /*$EPM_s\leftarrow EPM(s)\cup\{m\}$*/
\item \small \hspace{2mm} Refine($M$,$R$);
\item \small \hspace{2mm}
\emph{found}$\leftarrow$Node/EdgePathMappingSearch($s$, $M$, $R$);
\item \small \hspace{2mm} \textbf{if}(\emph{found})
\item \small \hspace{5mm} \textbf{return true};
\item \small \hspace{2mm} \textbf{else}
\item \small \hspace{5mm} $s\leftarrow$RecoverState($s'$);
\item \small \textbf{return false};
\end{enumerate}

From line 1-2, we can see that when a new state $s$ arrives, $s$ can
be a \emph{dead state} or \emph{success state}(complete mapping
state). The state space search arrives at a \emph{success state} if
all the node mappings or edge-path mappings have been found, which
means $|NM_s|=|V_{1}|$ or $|EPM_s|=|E_{1}|$, where $NM_s$ and
$EPM_s$ are the node match set and edge-path match set at state $s$.
The node mapping state space search arrives at a \emph{dead state}
if there is a row with all $0$'s in node compatible matrix $M$ of
the current state, i.e. $\exists i, |NM_s|\leq i \leq n_1$,s.t.
$\sum_{1\leq j \leq n_2}m_{ij}=0$. And the edge-path mapping state
space search arrives at a \emph{dead state} if there is no path
between any one branch node pairs, i.e., $\exists i, |EPM_s|\leq i
\leq n'_1$,s.t. $\prod_{(f^{-1}(node(i)),f^{-1}(node(j)))\in
E_1}r_{ij}=0$, where $node(i)$ gets the vertex corresponding to the
$i$th column in matrix $R$, which can be easily determined from
independent path matrix $R$ of the current state.

Any time the search process enters into a success state or dead
state, the procedure is over. If success state arrives, the complete
mapping is found and the procedure returns true. If dead state
arrives, the procedure returns false. On any other cases, the
procedure will continue exploring the state space.  The 1-11 lines
describe the process.

Assume the process comes to a state $s$ that is only a partial
solution. Then as long as there exists a valid mapping pair, i.e., a
node pair or an edge-path pair, we need to generate a new state by
adding the new match (line 9) to the existing solution $M_s$. To
enable backtracking, we need to backup the current state first (line
8), including the node compatible matrix, independent path matrix
etc. Because after a new match added to the current solution, these
two basic data structure will be refined to prune the following
mapping space (line 10). Then DFS continues, until the search enters
into dead state or success state. If we cannot find a success state
in subtree space rooted at $s$, we recover the state $s$(line 15),
and try the sibling state branches.

\subsection{Refinement Procedure}

To traverse all possible mapping branches is time consuming, so
space pruning is essential for ndSHD. For this purpose, we devise
two refinement procedures on $R$ and $M$, respectively, the
correctness of the former refinement is based on Lemma 3,4, and the
latter is based on Lemma 5.

\begin{lemma}
In the matching process, let $s$ be the current state, if $v\in
NM_s^{(2)}$ and $\mathcal{M}_s$ will be a partial solution of some
final solution $\mathcal{M}$, then any path with $v$ as inner vertex
will not $\in EPM^{(2)}$.
\end{lemma}

If $v\in NM_s^{(2)}$ and $NM_s^{(2)}$ will be a subset of some final
solution, then $v$ will a branch nodes of $G_{2}$. Since branch
nodes can only be the end vertex of the final independent path set,
thus any path with $v$ as inner vertex will not belong to the final
independent path set.

\begin{lemma}
In the matching process, let $s$ be the current state, if $p\in
EPM_s^{(2)}$ and $\mathcal{M}_s$ will be a partial solution of some
final solution $\mathcal{M}$, then any path passing trough the inner
vertex of $p$ will not $\in EPM^{(2)}$.
\end{lemma}

Obviously, if $p\in EPM_s^{(2)}$, then all the path passing through
any inner vertex of $p$ will joint with $p$, so all these paths will
not occur in $EPM^{(2)}$.

Lemma 3 implies that, if a vertex $v$ in $G_2$ is added to the
existing node match set, all the paths with $v$ as inner vertex can
be removed from RLists and the number in the corresponding element
in $R$ can be decreased. Lemma 4 implies that if we reach a new
state by adding a new edge-path pair $(e_{1i},p_{2i})$, all the path
passing through any inner vertex of $p_{2i}$ can be removed from
RLists and the number in the corresponding element in $R$ can be
reduced.

As shown in Figure \ref{fig:example}, if vertex $v_2$ in $G_1$ is
mapped to $v_8$ in $G_2$, any path passing through $v_8$ could be
removed from RList, thus the potential edge-path mapping space could
be pruned. As shown in Figure \ref{fig:run_example}, when $(13,296)$
is added to the partial solution, the subtree rooted at node
$(13-296)$ will be reduced, in the way that all the branches
containing paths passing through vertex $v_9$ will be pruned.

\begin{lemma}
In the matching process, let $s$ be the current state, if
$(v_{i},v_{j})\in NM(v_{i}\in V_{1},v_{j}\in V_{2})$ and
$\mathcal{M}_s$ will be a partial solution of some final solution
$\mathcal{M}$, then the following statements hold true:
\begin{enumerate}
\item $\prod r_{j'k}>0, where j'=Index(v_j), k\in Index(V)$ and $V=\{v_{2}|v_{1}\in Adjacent(v_{i}) \land (v_{1},v_{2})\in
NM_{s}\}$.
\item $\forall v'\in V',\exists v\in V_{2}$ such that
$l_{2}(v)=l_{1}(v')$ and $r_{j'k}>0, wher j'=Index(v_j),k=Index(v)$
and $V'=\{v'|v'\in Adjacent(v_i)\cap(V_1-NM_s^{(1)})\}$.
\item
The path set consisting of the paths to which all mentioned
$r_{j'k}$'s in (1) and (2) indicate is independent.
\end{enumerate}
\end{lemma}

In the above statements, the function $Index(v)$ gets an index in
$R$ for a node $v$ in $G_{2}$; and $Adjacent(v)$ obtains the
adjacent nodes set of $v$.

Suppose that the partial solution of current state will grow to be
one final successful solution, Lemma 5 implies that two node
$v_{i}\in V_1$ and $v_{j}\in V_2$ is compatible if only the three
conditions are satisfied, i.e., if any one is not satisfied,
$m_{ij}$ in node compatible matrix at state $s$, namely $M^{s}$,
could be refined to be 0.

 As shown in Figure \ref{fig:example}, assume current
 matching state is $s$, and $NM_s=\{(1,2),(2,8)\}$.
 Condition 1 implies that if (3,7) can be added to $NM$, i.e.,
$v_{3}$ in $G_{1}$ can be mapped to $v_{7}$ in $G_{2}$, there must
exist two independent paths from  $v_{7}$ to $v_{2}$ and $v_{8}$ in
$G_{2}$, otherwise $m_{37}$ can be refined to be $0$, thus the node
mapping space could be pruned. Moreover, since $v_3$ and $v_4$ are
adjacent in $G_1$, there must exist a corresponding path in $G_2$
from $v_{7}$ to some node with the same label as $v_{4}$ of $G_{1}$,
otherwise $m_{37}$ can be refined to be $0$, which is stated in
condition 2. Furthermore, all the above paths must be node disjoint,
which is implied in condition 3. Obviously, if $(l,h)$ is set as
(2,2), paths connecting $v_7$ to $v_2$,$v_4$ and $v_8$ all pass
through node $E$. Hence $m_{37}$ in matrix $M^{s}$ can be refined to
be $0$.

\subsection{More Efficient Searching Strategy}

A basic observation of the above refinement procedures is that the
constraint resulting from an edge-path match will be more restricted
than that resulting from a node match. Hence, a better strategy is
to try edge-path match as early as possible, instead of performing
edge-path match only after complete node match has been found. We
denote these two strategy as $s_1$ (old strategy) and $s_2$(new
strategy), respectively; and algorithms employing two strategies are
denoted as \emph{ndSHD1} and \emph{ndSHD2}, respectively.
Intuitively, in \emph{ndSHD2} the searching procedure will meet with
the dead state very early if the current searching path will not
lead to a successful mapping solution, thus the searching procedure
will fast backtrack to try another mapping solution.

 As an example, assume that $(l,h)$ is set as (2,2) and the current
matching state is $s$ such that $NM_s=\{(1,2),(2,8)\}$. Since $v_1$
and $v_2$ is adjacent in $G_1$, why we not try to match a path in
$G_2$ for the edge $e(v_1,v_2)$? If we do so, there are only two
suitable edge-path pairs $(12,298)$ and $(12,218)$. Then, once
$(12,298)$ has been added to $EPM_s$, all paths in $G_2$ passing
through $v_9$ will be excluded from $R$ and $Rlist$, thus the
searching space could be pruned early. Furthermore, we can see that
the current partial mapping solution will not be a part of a final
successful solution, thus any other solution with this partial
solution as subset will be pruned. And if we try the edge-path pair
$(12,218)$, we can eventually find a successful complete mapping
solution.

The framework of the  \emph{ndSHD2} is similar to that of
\emph{ndSHD1}, which is omitted here. The detailed procedure of
\emph{ndSHD2} is shown in NodeMappingSearch2 and
EdgePatbMappingSearch2. Note that, in NodeMappingSearch2, after a
new node match $(v_i,v_j)$ has been added to $NM_s$, we can get an
edge set consisting of edges that connect $v_i$ to any vertex
exiting in $NM_s^{(1)}$ (line 11), namely $E=\{(v_i,u)|u\in
NM_s^{(1)}\cap Adjacent(v_i)\}$. If $E=\emptyset$ (for connected
graph, it only happened at the initial stage of the first node
match), we continue the node mapping searching procedure (line
12-13); otherwise, we switch to edge-path mapping search procedure
to find valid path in $G_2$ for each edge in $E$ (line 14-15). In
EdgePatbMappingSearch2, if we can find valid paths for all the edges
in E, the edge-path mapping search procedure returns true (line 3-4)
and we will turn to the node mapping space search (line 13);
otherwise, we continue the edge-path mapping search procedure (the
while body).

\medskip
\noindent{\textbf{Algorithm} NodeMappingSearch2
($s$,$M$,$R$)\newline \noindent{\small \textbf{Input} and
\textbf{Output} is the same as that in NodeMappingSearch1.

\small\begin{enumerate}
\item \small \textbf{if}($s$ is \emph{complete mapping state})
\item \small \hspace{2mm} \textbf{return true};
\item \small \textbf{if}($s$ is \emph{dead state})
\item \small \hspace{2mm} \textbf{return false};
\item \small let \emph{found}$\leftarrow$\textbf{false}
\item \small \textbf{while}(\textbf{not} \emph{found} \&\& Exists
Valid node Mapping Pair)
\item \small \hspace{2mm} $m\leftarrow$GetNextNodePair(); /*Get a next valid node
pair*/
\item \small \hspace{2mm} $s'\leftarrow$BackupState($s$);
\item \small \hspace{2mm} $NM_s\leftarrow NM_s\cup\{m\}$;
\item \small \hspace{2mm} Refine($M$,$R$);
\item \small \hspace{2mm} $E\leftarrow$NewEdgeEmergent($s$,$G_1$)
\item \small \hspace{2mm} \textbf{if}$(E=\emptyset)$
\item \small \hspace{5mm} \emph{found}$\leftarrow$NodeMappingSearch2($s$, $M$, $R$);
\item \small \hspace{2mm} \textbf{else}
\item \small \hspace{5mm} \emph{found}$\leftarrow$EdgePathMappingSearch2($s$, $M$, $R$, $E$);
\item \small \hspace{2mm} \textbf{if}(\emph{found})
\item \small \hspace{5mm} \textbf{return true};
\item \small \hspace{2mm} \textbf{else}
\item \small \hspace{5mm} $s\leftarrow$RecoverState($s'$);
\item \small \hspace{2mm} \textbf{return} \emph{found};
\end{enumerate}

\medskip
\noindent{\textbf{Algorithm} EdgePathMappingSearch2
($s$,$M$,$R$,$E$)\newline \noindent{\small \textbf{Input} and
\textbf{Output} is the same as that in EdgeMappingSearch1 except
$E$, which is the edges in $G_1$ induced by $NM_s^{(1)}$.

\begin{enumerate}
\item \small \textbf{if}($s$ is \emph{dead state})
\item \small \hspace{2mm} \textbf{return false};
\item \small \textbf{if}($s$ is \emph{complete mapping state with respect to E})
\item \small \hspace{2mm} \textbf{return true};
\item \small let \emph{found}$\leftarrow$\textbf{false}
\item \small \textbf{while}(\textbf{not} \emph{found} \&\& Exists
Valid edge-path Mapping Pair)
\item \small \hspace{2mm} $m\leftarrow$GetNextEdgePathPair(); /*Get a next valid edge-path
pair*/
\item \small \hspace{2mm} $s'\leftarrow$BackupState($s$);
\item \small \hspace{2mm} $EPM_s\leftarrow EPM_s\cup\{m\}$;
\item \small \hspace{2mm} Refine($M$,$R$);
\item \small \hspace{2mm}
\emph{found}$\leftarrow$EdgePathMappingSearch2($s$, $M$, $R$, $E$);
\item \small \hspace{2mm} \textbf{if}(\emph{found})
\item \small \hspace{5mm} \emph{found}$\leftarrow$NodeMappingSearch2($s$, $M$, $R$);
\item \small \hspace{2mm} \textbf{else}
\item \small \hspace{5mm} $s\leftarrow$RecoverState($s'$);
\item \small \textbf{return false};
\end{enumerate}

\section{Experimental Evaluation}

To test the efficiency of the algorithm, we generate the synthetic
data sets according to the random graph \cite{ref8} model that links
each node pair by probability $p$. All generated graphs are vertex
labeled undirected connected graphs. We also randomly label every
node so that the vertex labels are uniformly distributed. We
implement the algorithm in C++, and carry out our experiments on a
Windows 2003 server machine with Intel 2GHz CPU and 1G main memory.

The efficiency of the algorithms is influenced by the following
factors: $N_{1}$: node size of $G_{1}$, $N_{2}$: node size of
$G_{2}$, $M_{1}$: average degree of $G_{1}$, $M_{2}$: average degree
of $G_{2}$, $(L,H)$: the minimal and maximal path length. The
efficiency also can be influenced by the number of vertex labels.
Obviously, large number of labels will exert great constraint on the
initial node compatible matrix, thus reduce the runtime
significantly. However, for determination algorithms, the runtime
also may be influenced by the answer to the determination.
Generally, if the answer is false, in the worst case the algorithm
may need to traverse the entire mapping spaces, which is very time
consuming. If the answer is true, then in the best case the
algorithm may only need to try one complete match procedure.
However, in the following experiments, we can see that result of the
determination has limited impact on the runtime, which could partly
be attributed to the strong pruning ability of the refinement
procedure. Due to these pruning techniques, even the result is
false, the procedure will backtrack as early as possible, thus the
whole runtime is rarely impacted.

\begin{figure}
\centering \subfigure[$ndSHD1$] { \label{fig:exp1:s1}
\includegraphics[scale=0.4]{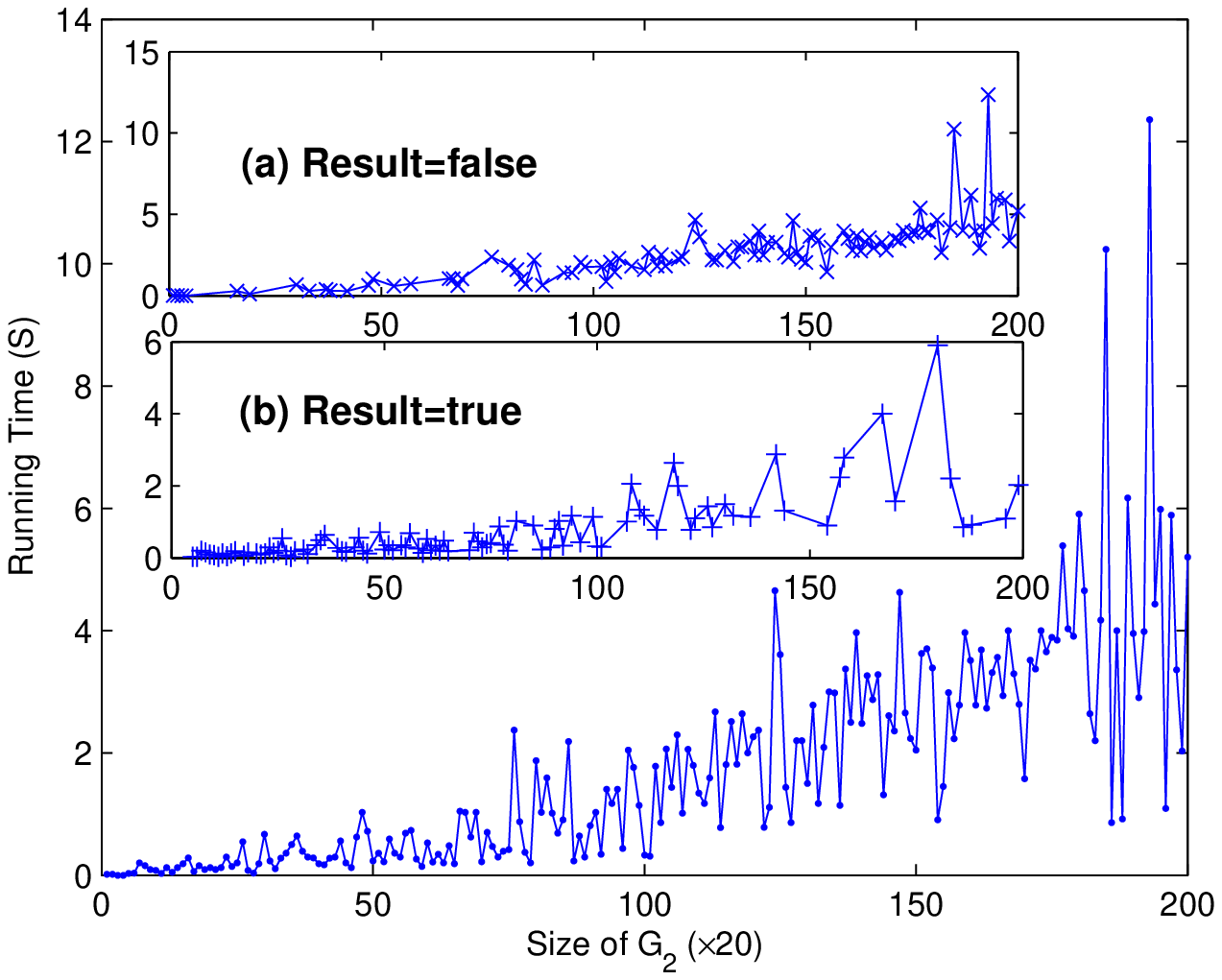}}
\subfigure[$ndSHD2$] { \label{fig:exp1:s2}
\includegraphics[scale=0.4]{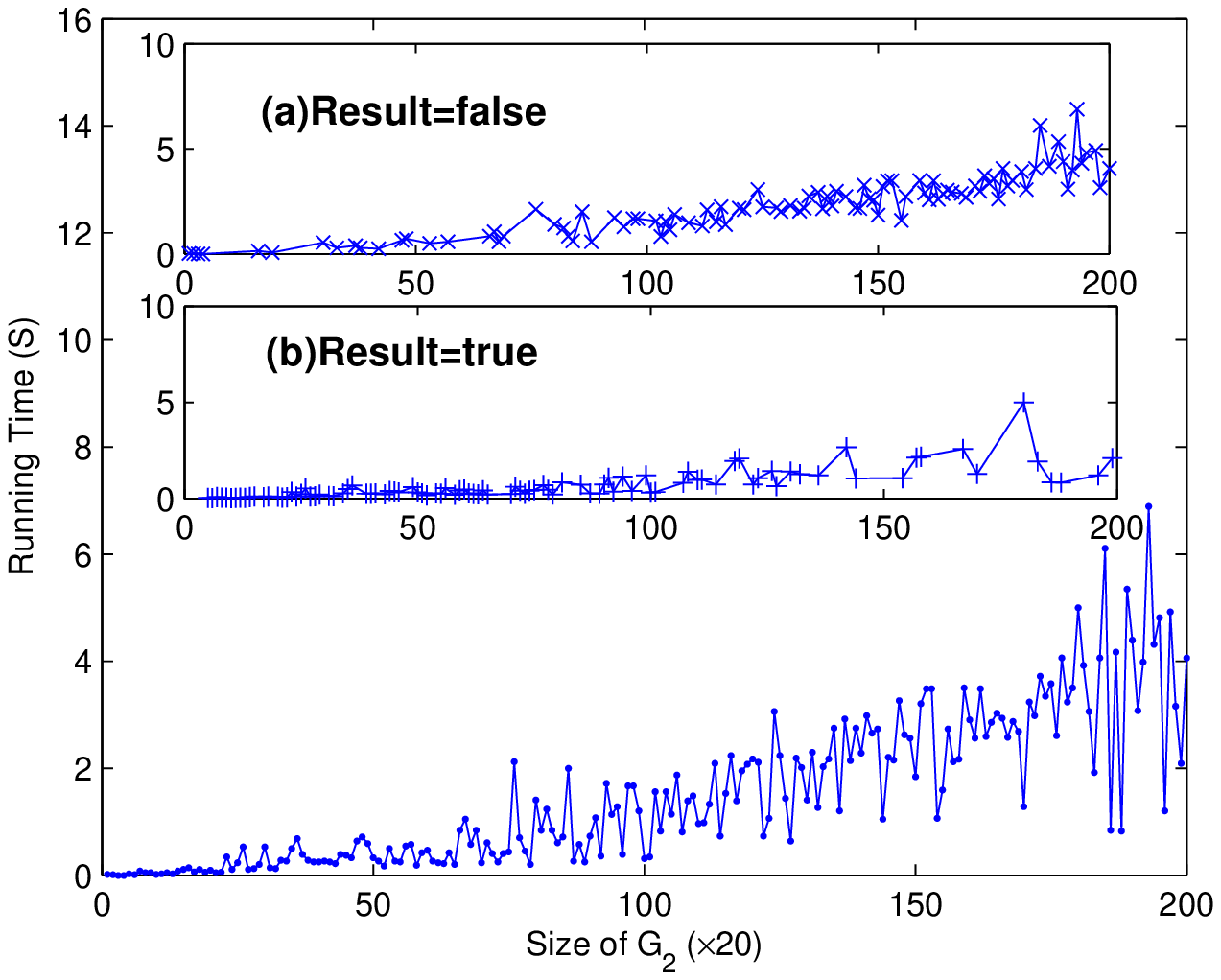}}
\caption{Efficiency and scalability with respect to the growth of
size of data graph ($G_2$). The inset of (a), (b) show runtime of
all running cases that the determination result is \emph{true},
\emph{false} respectively. }
\label{fig:exp1} %% label for entire figure
\end{figure}

First we will demonstrate the scalability with respect to the growth
of the size of nodes of data graph $G_2$ via experiment $Exp_1$. We
use a complete graph with 4 uniquely labeled nodes, denoted as
$C_4$, as a minor graph; we generate overall 200 data graphs $G_2$
with node size varying from $20$ to $4000$ in an increment of 20.
The average degree of each data graph is fixed as 4 and nodes of
each graph are randomly labeled as one of overall 20 labels. $L$ and
$H$ are fixed as 1 and 3, respectively, meaning that the path length
is in the range of $[1,3]$. Thus the parameters can be denoted as
$N_{1}4M_{1}3M_{2}4L1H3$. From Figure \ref{fig:exp1}, we can see
that \emph{ndSHD1} and \emph{ndSHD2} both are approximately linearly
scalable with respect to the number of nodes in $G_2$, irrespective
of the result of the determination result. Notice that for $G_{2}$
with about $4000$ nodes and $8000$ edges, the worst case of
\emph{ndSHD1}is no more than $13$s, the worst case of \emph{ndSHD2}
is no more than $7$s.

$Exp2$ is designed to show the scalability of \emph{ndSHD1} and
\emph{ndSHD2} with respect to the size of $G_1$, where we fix some
parameters as $M_{1}4N_24kL1H3$ and vary the size of $G_1$ from 6 to
82 in increment of 4 to generate 20 minor graphs. Each minor graph
is uniquely labeled, meaning that the number of labels equals to
that of nodes. Two data graphs are used, one has average degree
$M_2$ as 8 and the other as 20. These two data graphs are randomly
labeled as one of 200 labels. Figure \ref{fig:exp2}(a) and (b) show
the results with $M_2$ set as $8$ and $20$, respectively; and these
two experiments are denoted as $Exp2_1$ and $Exp2_2$, respectively.
The determination results of running case shown in Figure
\ref{fig:exp2}(a) are all false due to the relative sparsity of
$G_2$; and determination results of all running cases shown in
Figure\ref{fig:exp2}(b) are true due to relatively higher density of
$G_2$. As can be seen, \emph{ndSHD1} and \emph{ndSHD2} both are
approximately linearly scalable with respect to the number of nodes
in $G_1$. We also can see that, when $G_2$ is sparse, the difference
of performance between \emph{ndSHD1} and \emph{ndSHD2} are so minute
that can not be discerned; whereas as $G_2$ becomes denser, running
time of \emph{ndSHD1} is not available, meaning that all running
cases need time larger than one hour, while runtime increase of
\emph{ndSHD2} is not very substantial.

\begin{figure}
\begin{minipage}[t]{0.6\linewidth}
\centering
\includegraphics[width=2.74in]{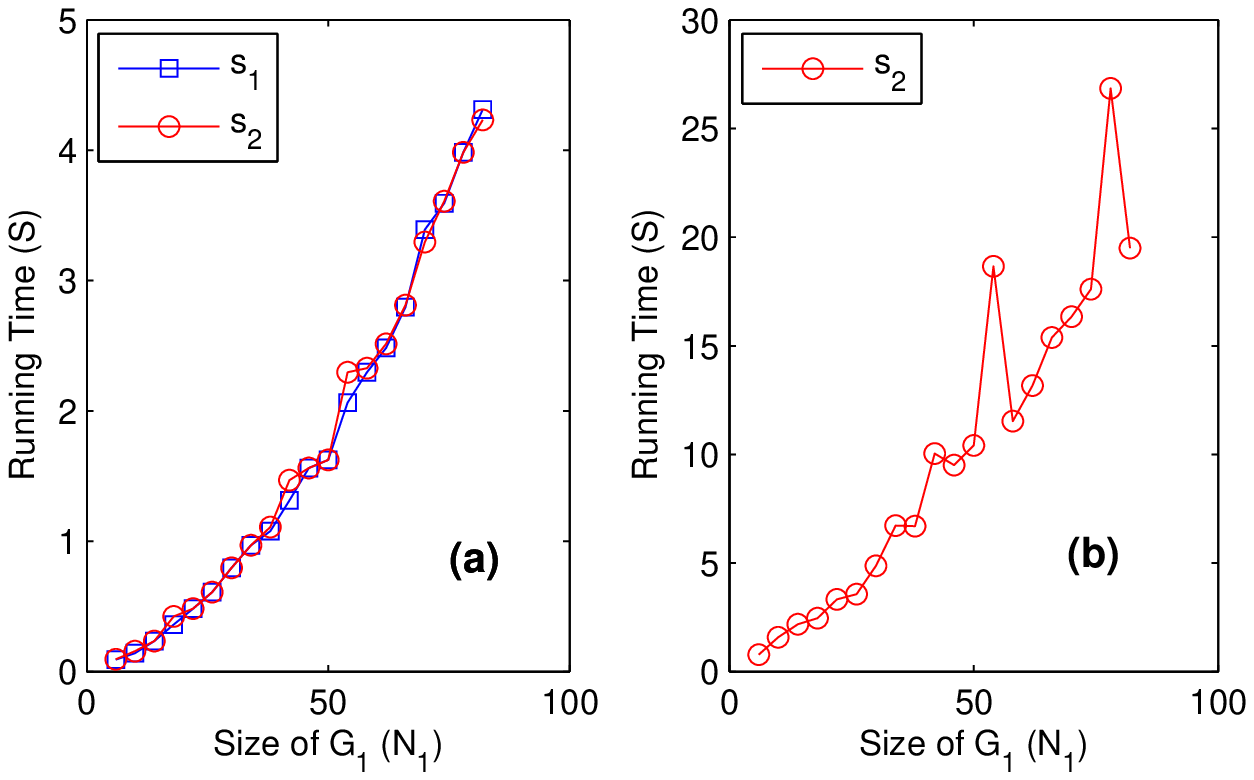}
\caption{Scalability with respect to the size of $G_1$.  }
\label{fig:exp2}
\end{minipage}
\begin{minipage}[t]{0.4\linewidth}
\centering
\includegraphics[width=2.08in]{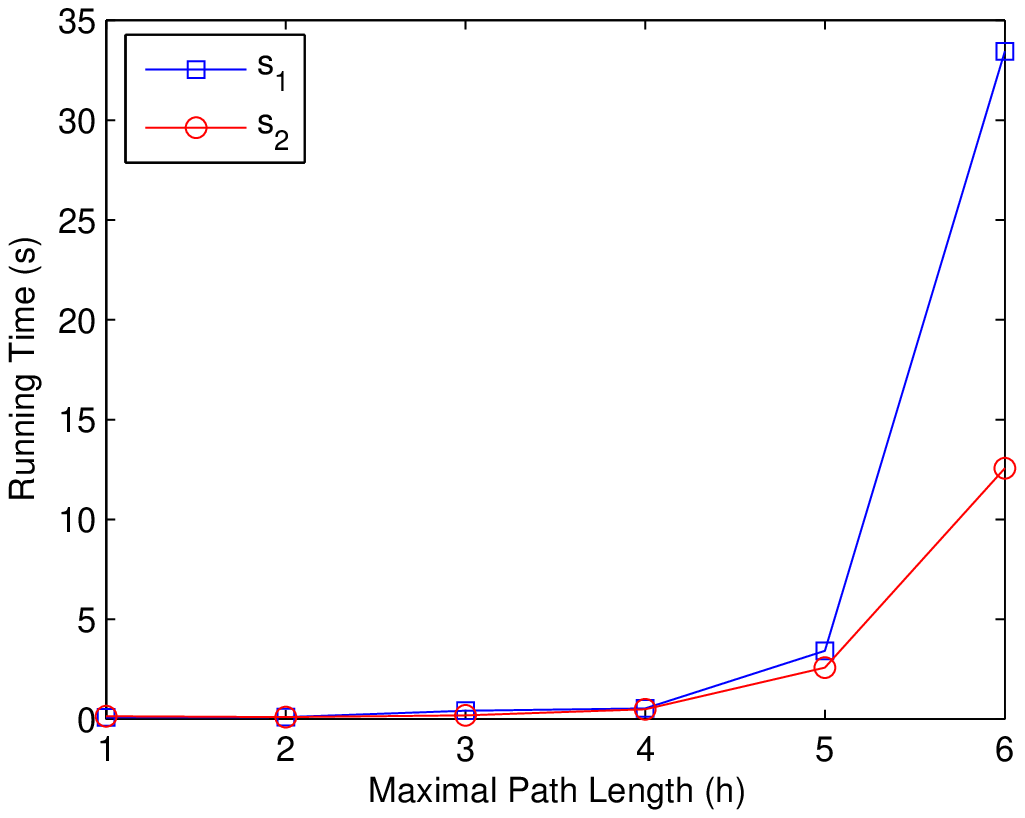}
\caption{Scalability with respect to the upper bound of path length}
\label{fig:exp4}
\end{minipage}
\end{figure}

$Exp3$ is designed to show the scalability with respect to the
growth of density of $G_2$, parameters are fixed as
$N_16M_{1}5N_21kL1H3$. Minor graph are uniquely labeled; data graphs
are randomly labeled as one of 20 labels. Table \ref{tab:exp3_data}
shows the running time when we vary $M_2$ from 2 to 20 in increment
of 1. As can be seen, runtime of \emph{ndSHD1} and \emph{ndSHD2}
both approximately increase linearly with the growth of $M_{2}$.
However, we must note that for \emph{ndSHD1}, there exists some
outliers which consume too much time, e.g, when $M_2=17$, more than
10 minutes are needed, when $M_2=16$ running time is not available.
Compared to \emph{ndSHD1}, \emph{ndSHD2} is more stable.

\begin{table}
%\begin{ruledtabular}
\centering \caption{\label{tab:exp3_data} Running time of $Exp3$}
\begin{tabular}{c|cccccccccc}
\hline%
\hline%
$M_{2}$ & 2 & 3 & 4 & 5 & 6 & 7 & 8 & 9 & 10& 11
 \\
\hline%
$Result$ & 0 & 0 & 0 & 0 & 1 & 1 & 1 & 1 & 1 & 1  \\
\hline%
$s_{1}$ & 0 & 0.14 & 0.31 & 1.78 & 21.67 & 19.69 & 1.77 & 1.23 &
83.66 & 661.33  \\
\hline%
$s_{2}$ & 0.016 & 0.14 & 0.33 & 1.05 & 11.41 & 0.97 & 0.92 & 1.27 &
2.14 & 3.77 \\
\hline%
\hline%
$M_{2}$& 12 & 13 & 14 & 15 & 16 & 17 & 18 & 19 & 20\\
\hline%
$Result$ & 1 & 1 & 1 & 1 & 1 & 1
& 1 & 1 & 1\\
\hline%
$s_{1}$& 3.23 & 3.89 & 4.24 & 5.36 & - & 6508.7 &
7.75 & 10.06 & 9.77\\
\hline%
$s_{2}$& 3.11 & 3.92 & 4.23 & 5.36 & 6.66 & 7.91
& 8.06 & 10.72 & 10.11 \\
\hline%
\hline%
\end{tabular}
\end{table}

Figure \ref{fig:exp4} shows the runtime of the algorithm with
respect to $(l,h)$. Parameters of this experiment (denoted as
$Exp4$), are set as $N_{1}4M_{1}3N_21kM_{2}8L1$. The vertex labeling
of minor graph and data graphs are the same as $Exp3$. As can been
seen the broader the range is, the longer the running time is; and
the runtime of \emph{ndSHD1} and \emph{ndSHD2} both increase
dramatically with the growth of upper bound of path length. However,
the increasing speed of \emph{ndSHD2} is slower than that of
\emph{ndSHD1}, which implies that \emph{ndSHD2} is more efficient
than \emph{ndSHD1} with respect to larger $h$. The super linearly
growth of the runtime with the increase of upper bound of the path
length can be partly attributed to the exponentially growth of the
number of potential mapped paths. Luckily, in the real applications,
larger upper bound is too unrestricted when performing fuzzy
matching on graph data, thus usually upper bounds less than 3 are
used.

\begin{figure}
\centering
\includegraphics[width=2in]{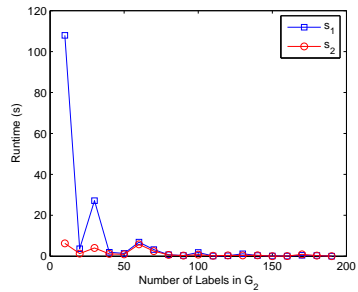}
\caption{Effect of label numbers of $G_2$ on the performance of
\emph{ndSHD1} and \emph{ndSHD2}.} \label{fig:exp5}
\end{figure}

To examine the impact of number of vertex labels on the performance
of \emph{ndSHD1} and \emph{ndSHD2}, we use a uniquely labeled graph
with 6 nodes and 15 edges as minor graph, a graph with 1000 nodes
and 4000 edges as data graph. We randomly labeled the data graph
from 10 labels to 200 labels in increment of 10 to generate 20
different labeled data graphs. $L$ and $H$ are set as 1 and 3,
respectively. The result of this experiment (denoted as $Exp5$) is
shown in Figure \ref{fig:exp5}. Clearly, runtime of \emph{ndSHD1}
and \emph{ndSHD2} substantially decrease with the growth of number
of labels of $G_2$, which confirms to what we have expected, since
larger number of labels in $G_2$ can reduce the node mapping space
between minor graph and data graph. We also can see that
\emph{ndSHD2} outperforms \emph{ndSHD1} to a great extent when label
number is small.

To examine the stability of \emph{ndSHD1} and \emph{ndSHD2}, we
recorded in Table \ref{tab:stability} the statistics including
\emph{max}, \emph{mean} and \emph{standard deviation} of sample data
used in the above experiments. We can see that in almost all the
experiments, the standard deviation of \emph{ndSHD2} is much less
than that of \emph{ndSHD1}, indicating that \emph{ndSHD2} is more
stable than \emph{ndSHD1}.

In a summary, in some simple running cases, such as small size of
minor graph, sparse data graphs, small value of upper bound of path
length, larger number of labels in data graphs, both \emph{ndSHD2}
and \emph{ndSHD1} are scalable and efficient. However, in more
complex cases, \emph{ndSHD2} will outperform \emph{ndSHD1}
substantially in all aspects, including scalability, efficiency and
stability.

In Figure \ref{fig:exp6},we also illustrate the detailed searching
procedure of two running cases to show the superiority of
\emph{ndSHD2} to \emph{ndSHD1}. In both of these two cases,
\emph{ndSHD2} runs much faster than \emph{ndSHD1}. Since the mapping
searching procedure has been designed to be a recursive procedure,
statistics about the recursive depth of each match (node-node match
or edge-path match) will be a significant index indicating the
performance of the algorithm. Hence, we recorded recursive depth of
all matches in the searching procedure. Obviously, either narrow
width of the exploring space or small value of the average backtrack
depth, will lead to the less runtime of the algorithm. Hence ,from
Figure \ref{fig:exp6}, we can easily see the great advantage of
\emph{ndSHD2} over \emph{ndSHD1}, which can be attributed to the
small value of the width or average backtrack depth in the actual
exploring space.

\begin{figure}
\centering
\includegraphics[width=3.5in]{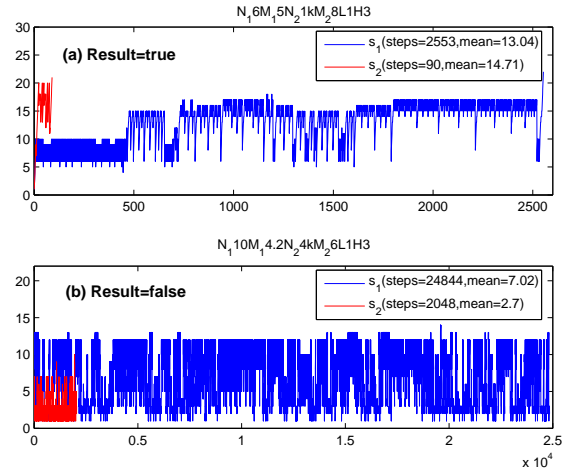}
\caption{Recursive depths of two running cases } \label{fig:exp6}
\end{figure}

\begin{table}
\centering \caption{\label{tab:stability}Statistics of sample data
in 5 experiments }
\begin{tabular}{c cc cc cc cc cc cc}
\hline  &\multicolumn{2}{c}{$Exp1$\text{  }}
&\multicolumn{2}{c}{$Exp2_1$\text{  }}
&\multicolumn{2}{c}{$Exp2_2$\text{  }}
&\multicolumn{2}{c}{$Exp3$\text{  }}
&\multicolumn{2}{c}{$Exp4$\text{  }}
&\multicolumn{2}{c}{$Exp5$\text{  }}\\
statistics&\text{ }$s_1$&\text{ }$s_2$&  \text{   }$s_1$&\text{ }$s_2$&  \text{   }$s_1$&\text{ }$s_2$&  \text{   }$s_1$&\text{ }$s_2$&  \text{   }$s_1$& \text{ }$s_2$ &\text{   }$s_1$&\text{ }$s_2$\\
\hline
max& 12.36 &6.89 & 4.312 & 4.234 &-& 26.84 &6509 & 11.41 &33.44 &12.56 &107.9 &6.297\\
mean& 1.727&1.49 & 1.709 & 1.73  &-& 10.06 &408  & 4.32  &6.325 &2.675 &8.312 &1.392\\
std&1.788&1.396  & 1.344 & 1.331 &-& 7.24  &1530 & 3.797 &13.34 &4.937 &24.89 &1.909\\
 \hline
\end{tabular}
\end{table}

\section{Conclusions}

In this paper, we investigated the problem known as node disjoint
subgraph homeomorphism determination; and proposed two practical
algorithms to address this problem, where many efficient heuristics
have been exploited to prune the futile searching space. The
experimental results on synthetic data sets show that our algorithms
are scalable and efficient. To the best of our knowledge, no
practical algorithm is available to solve node disjoint subgraph
homeomorphism determination.

\end{document}